# Interplay between topological protected valley and quantum Hall edge transport

**Authors:** Fabian R. Geisenhof[1], Felix Winterer[1], Anna M. Seiler[1,2], Jakob Lenz[1], Ivar Martin[3], R. Thomas Weitz[1,2,4,5]*

**Affiliations:**

[1]Physics of Nanosystems, Department of Physics, Ludwig-Maximilians-Universität München, Geschwister-Scholl-Platz 1, Munich 80539, Germany

[2]1st Physical Institute, Faculty of Physics, University of Göttingen, Friedrich-Hund-Platz 1, Göttingen 37077, Germany

[3]Materials Science Division, Argonne National Laboratory, Lemont, Illinois 60439, USA

[4]Center for Nanoscience (CeNS), LMU Munich, Schellingstrasse 4, Munich 80799, Germany

[5]Munich Center for Quantum Science and Technology (MCQST), Schellingstrasse 4, Munich 80799, Germany

*Corresponding author. Email: thomas.weitz@uni-goettingen.de

# Abstract:

An established way of realizing topologically protected states in a two-dimensional electron gas is by applying a perpendicular magnetic field thus creating quantum Hall edge channels. In electrostatically gapped bilayer graphene intriguingly, even in the absence of a magnetic field topologically protected electronic states can emerge at naturally occurring stacking domain walls. While individually both types of topologically protected states have been investigated, their intriguing interplay remains poorly understood. Here, we focus on the interplay between topological domain wall states and quantum Hall edge transport within the eight-fold degenerate zeroth Landau level of high-quality suspended bilayer graphene. We find that the two-terminal conductance remains approximately constant for low magnetic fields throughout the distinct quantum Hall states since the conduction channels are traded between domain wall and device edges. For high magnetic fields, however, we observe evidence of transport suppression at the domain wall, which can be attributed to the emergence of spectral minigaps. This indicates that stacking domain walls do potentially do not correspond to a topological domain wall in the order parameter.

# Introduction:

Electrons near the Fermi surface of two-dimensional hexagonal materials typically occupy two or more distinct electronic valleys. The valley index adds to the carrier's charge and spin, enabling additional channels for spontaneous symmetry breaking at low temperatures, whereby valleys are polarised independently or in combination with charge and spin degrees of freedom[1,2]. The most direct way to induce non-trivial valley response is by breaking sublattice symmetry. This occurs naturally in boron nitride, which makes it a quantum Valley Hall insulator[3]. In Bernal-stacked bilayer graphene, the same effect is achieved by applying an interlayer bias[4]. Moreover, by spatially varying its sign, topological domain walls can be created, which exhibit one-dimensional (1D) electronic channels with quantized conductance[4], resilient to backscattering[5]. These electronic domain wall states provide a flexible platform to study 1D transport[6–8] and correlated physics[9–11]. However, creating them by electrostatic gating is technically challenging. Fortunately, similar physics transpire at stacking domain walls (DWs) in bilayer graphene, where the stacking arrangement of graphene layers changes from AB to BA[12]. Such domain walls are common in naturally Bernal-stacked bilayer graphene[13–15] and even ubiquitous in twisted bilayer graphene[16,17], which is known for hosting superconductivity at a certain twist angle[18]. When a *uniform* electric field is applied to a bilayer graphene flake with a DW, topologically protected valley-helical states emerge along the dislocation, surrounded by insulating bulk[12,14,19]. Critically for the present work, stacking domain walls can have much richer interplay with spontaneous symmetry breaking in bilayer graphene[20–27] compared to artificially created ones, as not being forced by applied bias to have charge imbalance between layers. The interplay between stacking domain walls and spontaneous symmetry breaking is of peculiar interest in the presence of a quantizing magnetic field, since bilayer graphene exhibits a very rich phase diagram owing to the eight-fold degeneracy of the zero-energy Landau levels[28–30] (coming from two valleys, two orbital Landau level indices, and two spins – neglecting Zeeman splitting). Interactions lift the degeneracy by generating orderings, leading to quantum Hall plateaus at all integer filling fractions between –4 and 4[24,28–32]. This complex and intriguing regime shows a large variety of ways the internal symmetry can break spontaneously in the absence of externally induced layer polarisation. Within this manifold, the valley, sublattice, and layer index are rigidly locked. Since at the stacking domain wall the roles of the layers are exchanged, any ordering that is not a valley singlet is guaranteed to be affected. The goal of our work is to study this interplay by means of transport measurements. It cannot be fully explored in the artificial electrostatic domain walls as a matter of principle. We chose freestanding dually gated bilayer graphene devices as an ideal and versatile platform, since – as indicated by our measurements below – DWs remain stable during processing and suspension, however, it is still unclear whether they consistently survive the assembling of bilayer graphene/hexagonal boron nitride heterostructures.

## Results:

**Topological valley transport in the presence of an electric field induced gap**

At first, suitable bilayer graphene flakes were preselected using optical microscopy and subsequently investigated with scattering scanning near field microscopy[14,15,33]. Even though flakes show a smooth surface in the topography (Fig. 1a), the near-field amplitude image (Fig. 1b) can reveal stacking domain walls. Second, contacts were designed in two different configurations, as schematically illustrated in Fig. 1c. Either a DW was contacted on both ends (the DW goes along the channel separating two distinct domains, one with AB and one with BA stacking), or, alternatively, no domain wall was within the channel. Two devices are discussed exemplarily in the following: D1-DW of the former and D2 (which has been also investigated in ref.[27]) of the latter type. Data from additional domain-wall containing devices are shown in the Supplementary Information.

Using the dual-gate structure and sweeping the top $V_t$ and bottom voltage $V_b$ while tracking the resistance for the two configurations reveals differences in their signatures (Fig. 1d,e). Device D2 (Fig. 1e) shows, consistent with previous measurements, the spontaneously gapped state at the charge neutrality point[20–24] and a phase transition to the insulating fully layer polarised state for increasing electric fields[23,24]. The resistance in device D1-DW (Fig. 1d) shows an overall similar behaviour, but, with very different values. This becomes more apparent when examining line traces (see Fig. 1f,g). Although the resistance in both devices behaves non-monotonically as a function of increasing $V_t$, which indicates the emergence of the layer antiferromagnetic (LAF) ground state with opposite spins in two layers[1,34,35] at charge neutrality and zero electric field (at $V_t \approx V_b \approx 0$), it remains low in device D1-DW. As discussed below, this is caused by additional charge channels, which mask the insulting phase. Moreover, consistent with previous measurements[7,14], the resistance saturates for an increasing electric field (here at $R \approx 8.5 \, k\Omega$), which unambiguously demonstrates the presence of zero-energy line modes[4,12,19]. In other words, although the perpendicular applied electric field induces a band gap within the system[36], topologically protected states at the $K/K'$ valleys persist, giving rise to helical valley transport (see the insets of Fig. 1d,e). Using the Landau-Büttiker formula[14] $\sigma = \sigma_0 \left(1 + \frac{L}{\lambda_m}\right)^{-1}$ with a channel length of $L = 0.7 \, \mu m$ and the theoretical conductance of the domain wall of $\sigma_0 = 4 \, e^2/h$ (where $e$ is the electronic charge and $h$ Planck's constant) results in a mean free path of $\lambda_m \approx 2.2 \, \mu m$. With $\lambda_m > L$, ballistic charge transport supported by the domain wall is confirmed, highlighting the high quality of the device[8,14].

**Behaviour of the kink states in the presence of broken-symmetry phases at low magnetic field**

Whereas artificially constructed domain walls can only be investigated in the presence of a perpendicular electric field[4,7,8] in a limited range of electric fields and densities, quantum transport along stacking domain walls have mostly been studied in zero magnetic field[14]. Hence, we focus here on the interplay of topologically domain walls and quantum Hall edge transport. Fig. 2a,b shows the conductance in the devices D1-DW and D2 as a function of charge carrier density $n$ and electric field $E$ at a magnetic field of $B = 3$ T. In both devices, the broken-symmetry states within the lowest Landau level octet[24,28–31] appear, however, with very different conductance values (see Fig. 2c). The emerging quantum Hall states in device D1-DW, although exhibiting unusual conductance values, can unambiguously be identified by examining their slope in fan diagrams (see Supplementary Fig. S1). Thus, the stacking domain wall in device D1-DW contributes additional charge transport channels in parallel to the quantum Hall edge states altering the overall conductance of the device. In fact, tracking the conductance of both devices as a function of density (Fig. 2c) reveals a conductance offset for most of the appearing states. In device D1-DW, the $\nu = 0$ state, which has previously been identified as an insulating canted antiferromagnetic (CAF) state[37,38], shows a rather high conductance of $\sigma \approx 2.9 \ e^2/h$ (see Fig. 2d). CAF states have been observed to have low edge conductance, attributed to the opening of a spectral minigap at the sample edges[2,37,38]. The observed high conductance is thus consistent with the maximum possible — four — kink states at the DW contributing to the charge transport (with a finite $\lambda_m \approx 1.9 \ \mu$m), as is also the case in the layer polarised (LP) $\nu = 0$ phase (see Supplementary Fig. S2 for more details) and in absence of $B$. For an increasing filling factor, the conductance changes to $\sigma \approx 3.5, 4.0$ and $3.9 \ e^2/h$ for the $\nu = -1, -2$ and $-4$ states (see Fig. 2d), respectively. This near constancy of conductance can be naturally explained: In the simplest model (see Fig. 2e), ignoring spin and orbital index for clarity, changing the Fermi level for an applied electrical field leads to the topological domain wall channels being traded for quantum Hall edge channels. Changing the filling factor from the electron to the hole side, exchanges the positions of the valley polarised channels. More precisely (see Fig. 2f), when increasing the filling factor, a domain wall channel disappears whereas an additional quantum Hall edge channel emerges. Hence, the conductance follows $\sigma = (4 - |\nu|)\sigma_{DW} + |\nu|\sigma_{QH}$ for $|\nu| \leq 4$, where $\sigma_{DW}$ is the conductance supported by a single kink state, and $\sigma_{QH} = e^2/h$ is the conductance of a quantum Hall edge channel. A linear fit to the data further supports this hypothesis (see Fig. 2d): for D2, it shows the expected slope of 1.0 $e^2/h$ per filling factor as there are only quantum Hall edge states present. On the contrary, it yields a slope of 0.23 $e^2/h$ per filling factor for device D1-DW. Although in all states with $|\nu| \leq 4$ four quantized channels contribute to the charge transport in total, the non-zero slope corresponds to the difference in conductance of the kink and edge states and shows that for increasing filling factor kink states with a lower conductance $\sigma_{DW} \approx$ 0.77 $e^2/h$ are traded for quantum Hall edge states with $\sigma_{QH} = e^2/h$. Consequently, the $\nu = \pm 4$ states seem to be free of the influence of the domain wall (see Fig. 2f). A more detailed consideration of the band structure reveals that stacking domain walls can affect even the higher Landau levels, albeit more weakly (see Supplementary Fig. S3). In our freestanding devices, these states are outside the accessible density regime at higher magnetic fields needed to observe the quantum Hall states.

**Emergence of a spectral minigap for high magnetic fields**

A more in-depth understanding of the intricate interplay between the quantum Hall edge modes and domain walls can be gained by investigating the charge transport at varying magnetic fields (see Fig. 3 and Supplementary Fig. S4 for more data). Line traces of the conductance as a function of filling factor measured in device D1-DW at zero and finite electric field show the $\nu = 0, \pm1, \pm2$ states (see Fig. 3a). In addition, we plot the conductance as a function of magnetic field for the individual states shown in Fig. 3b. Note that the conductance was averaged over the electric field range at which the respective state emerges, i.e. for the $\nu = 0$ CAF phase around zero electric fields, for the $\nu = -1$ and $-2$ at $|E| \geq 10$ mV/nm and $|E| \geq 15$ mV/nm, respectively, and for the $\nu = -4$ state at all electric fields.

Most prominently, we see a sharp dip to very low conductance around zero charge carrier density within the $\nu = 0$ phase at high magnetic fields of $B \geq 8$ T (marked with a cross in Fig. 3a), which can also be tracked as function of magnetic field (see Fig. 3b). Towards $B = 0$, the $\nu = 0$ state corresponds to the layer antiferromagnetic phase with spin and valley indices locked[1,34,35]. In general, we find high conductance in this regime, suggesting the presence of zero-energy line modes at the kink. This observation would be consistent with the LAF order parameter experiencing an order parameter reversal as illustrated in Fig. 3c. The 1D modes persist within the gap because counter-propagating states in the same valley have opposite spin, and hence scattering is suppressed. However, as the magnetic field is increased, spins cant and the LAF phase evolves into the canted antiferromagnetic phase[37,38]. Then, the counterpropagating modes in the same valley become partially spin aligned and can hybridize causing the emergence of a minigap. This is similar to the effect at the device edge. However, in the latter case the termination and backscattering off atomic scale defects can also couple opposite valleys[5], leading to further suppression of conductance. Our experimental data are indeed consistent with the opening of a gap and – when the Fermi level is located in this gap – a decrease in conductance. Outside of the gap, we expect a finite conductance, with a value determined by a sequence of the crossing bands and gap openings (see Fig 3c). Since canting of spins gets stronger with magnetic field, one can expect the size of the minigap to grow with increasing $B$. This is consistent with our experimental observations of decreasing conductance (see Fig. 3a,b and Supplementary Fig. S4) and could be the reason why we can only resolve the minigap at $B \geq 8$ T. Eventually, for an infinite perpendicular or a finite in-plane magnetic field the CAF phase is expected to evolve into the ferromagnetic phase[37,38], in which the stacking domain wall has little or no effect on the Landau level energy (see Fig. 3c), making the stacking domain wall effectively invisible (this regime was not investigated experimentally in this study).

Notably, the conductance of the $\nu = \pm 1, \pm 2$ states also decrease with increasing $B$ (Fig. 3a,b), whereas device D2 shows the expected constant values as a function of $B$ for each QH state (see Fig. 3b). These quantum Hall states occur in sufficiently large electric field states, and thus the valley polarisation is expected to change sign across the domain wall. In contrast, the spin polarisation remains constant across domain wall, pinned to the direction of magnetic field (see Fig. 2f). Therefore, the counterpropagating states at the domain wall belong to opposite valleys but same spin and can only be destroyed by local defects that can provide large momentum scattering. That is in contrast to the CAF states at $\nu = 0$ and $E = 0$, where a minigap can open owing to the hybridization of states within the same valley and without the need for short range scattering. The measurements indicate that increasing the magnetic field increases the intervalley scattering, although the exact mechanism at this point remains unclear. One possible explanation could be the change in relative spatial arrangement of the counterpropagating channels as a function of magnetic field (see Fig. 3d). Clearly, increasing the channel separation should suppress backscattering, and vice versa. An effect of this type has been observed in artificial domain wall states, where application of magnetic field or change of the chemical potential was found to affect the domain wall conductance[7]. Another possibility could be the that increasing magnetic field pushes the system towards new broken-symmetry states[39,40], which would change the order parameter and hence the behaviour of the kink states. However, these states have been observed only at very high magnetic fields and since we see no evidence of phase transitions in sample D2 for the same parameters, this possibility appears unlikely. Given that the measurements were performed in a two-terminal configuration, one should also make sure that the effect that we observe is not a consequence of a magnetic field dependent contact resistance of the kink states. However, we do not observe this behaviour for quantum Hall edge states (see Fig. 3b), and it is likely that the contact resistance of both types of one-dimensional channels behaves similarly. Additional devices revealed similar behaviours of the domain wall conductance with increasing magnetic field (see Supplementary Fig. S5).

**Temperature dependence of the domain wall states**

As final investigation to establish the interplay between edge and domain walls, we have conducted temperature dependent measurements. In Fig. 4, the conductance is shown as a function of temperature measured in different phases: in the layer antiferromagnetic, the canted antiferromagnetic as well as the layer polarised $\nu = 0$ phases and in the $\nu = -4$ phase. In contrast to device D2, which shows an activated temperature dependence of the conductance in all phases, D1-DW exhibits a much weaker temperature dependence and, most importantly, a finite conductance at low temperatures for the insulating LAF, CAF and LP phases (see Fig. 4a – c, respectively). As the charge channels induced by the stacking domain wall contribute in parallel to any edge channels, we can subtract the data measured in both devices to reveal the underlying temperature dependence of the domain wall $\sigma_{DW}(T) \approx \sigma_{diff}(T) = \sigma_{D1-DW}(T) - \sigma_{D2}(T)$, assuming that the activated charge transport behaves similarly in both devices. Notably, in all three $\nu = 0$ phases the difference $\sigma_{diff}(T)$ shows an approximately constant behaviour at low temperatures with $\sigma_{diff} \approx 2.5 - 3.5 \ e^2/h$ and only a slight increase in the LAF and CAF phases for $T \geq 3$ K. Overall, this weak temperature dependence is expected for 1D charge transport and suggests weakly localized metallic behaviour[41]. On the contrary, the $\nu = -4$ phase (see Fig. 4d) shows the same activated temperature dependence and very similar conductance values in both devices, indicating that the domain wall has negligible influence on the quantum transport in this phase.

**Conclusion and outlook:**

In conclusion, we have investigated the impact of stacking domain walls on the eight-fold degenerate zero energy Landau level in bilayer graphene. For future measurements, unexplored aspects such as the behaviour of domain walls within the $\nu = 0$ ferromagnetic phase at high in-plane magnetic fields as well as their impact on the energy landscape of higher Landau levels would be interesting to investigate. Furthermore, having established that in the lowest Landau level the edge states and domain wall channels co-exist, one can imagine investigating their mutual interaction[42] in narrow samples.

## Methods:

Bilayer graphene was exfoliated from a highly ordered pyrolytic graphite (HOPG) block onto Si/SiO$_2$ substrates and suitable flakes were preselected using optical microscopy. Afterwards, infrared nano-imaging[43] was performed in a scattering-type scanning near-field microscope (s-SNOM, neaspec GmbH) in tapping mode to detect any stacking domain walls. Hereby, an infrared CO$_2$ laser beam (with a wavelength of 10.5 µm) was focused onto a metal-coated atomic force microscopy tips (Pt/Ir, Arrow NCPT-50, Nanoworld), which was oscillating with a frequency and amplitude of 250 – 270 kHz and 50 – 80 nm, respectively. With this method, we were able to obtain topographic and infrared nano-images simultaneously. Electrodes (Cr/Au, 5/100 nm) in two distinct configurations, a top gate (Cr/Au, 5/160 nm) as well as a spacer (SiO$_2$, 140 nm) were fabricated using several steps of standard lithography techniques and electron beam evaporation. Subsequently, the devices were submersed in hydrofluoric acid to etch about 150 – 200 nm of the SiO$_2$ and consequently suspend both the top gates and bilayer graphene flakes. After loading the freestanding dually gated bilayer graphene devices into a dilution refrigerator current annealing was performed at 1.6 K. In devices without domain wall best results were obtained when using a current of about 0.35 mA/μm per layer. In devices with domain wall 150 – 250 % more current was needed to achieve a current saturation due to their lower resistance and shorter channels. All quantum transport measurements were conducted at the base temperature of the cryostat ($T$ < 10 mK), if not noted differently. Moreover, an excitation a.c. bias current of 0.1 – 10 nA at 78 Hz and Stanford Research Systems SR865A and SR830 lock-in amplifiers were used for the measurements, as well as Keithley 2450 SourceMeters to apply the gate voltages. Low-pass filters were used in series to reduce high frequency noise.

**Acknowledgement:** R.T.W. and F.R.G. acknowledge funding from the Center for Nanoscience (CeNS) and by the Deutsche Forschungsgemeinschaft (DFG, German Research Foundation) under Germany's Excellence Strategy-EXC-2111-390814868 (MCQST). I.M. was supported by the Materials Sciences and Engineering Division, Basic Energy Sciences, Office of Science, U.S. Dept. of Energy. We also thank Y.C. Durmaz and F. Keilmann for experimental assistance with the near-field optical microscopy.

**Author contributions:** F.R.G. fabricated the devices and conducted the measurements and data analysis. I.M contributed the theoretical part. All authors discussed and interpreted the data. R.T.W. supervised the experiments and the analysis. The manuscript was prepared by F.R.G., I.M. and R.T.W with input from all authors.

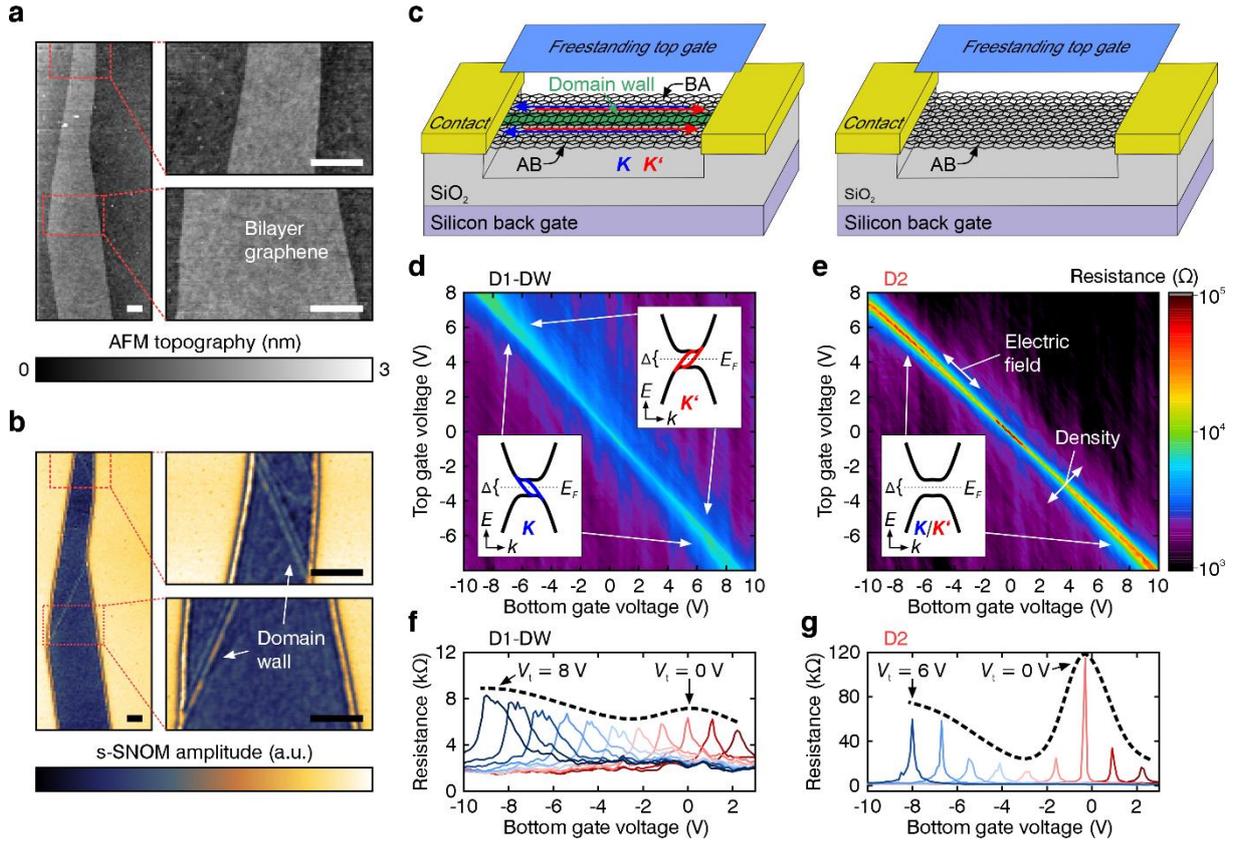

**Fig. 1 | Topologically protected states in bilayer graphene. a,b,** Atomic force microscopy image (a) and scattering-type scanning near-field microscopy image (b) of a bilayer graphene flake, with high-resolution zoom-in scans on the right. The scale bars are 0.5 μm. **c,** Freestanding dually gated bilayer graphene devices schematically shown with (left) and without domain wall (right) connecting the contacts. Topological valley transport along the domain wall is shown in blue and red in the $K$- and $K'$-valley, respectively. **d,e,** Resistance map as a function of top and bottom gate voltage for device D1-DW (d, with domain wall) and D2 (e, without domain wall). Insets: Electronic band structure of bilayer graphene with (d) and without a domain wall (e) for an applied electric field. Δ is the electric field induced bandgap, $E_F$ the Fermi level and the blue (red) lines indicate topologically protected, doubly spin degenerate chiral states in the $K(K')$-valley. **f,g,** Trace of the resistance as a function of $V_b$ for various $V_t$ with steps of 1 V shown for device D1-DW (f) and D2 (g). The dashed lines indicate the envelope of the resistance and are a guide to the eye.

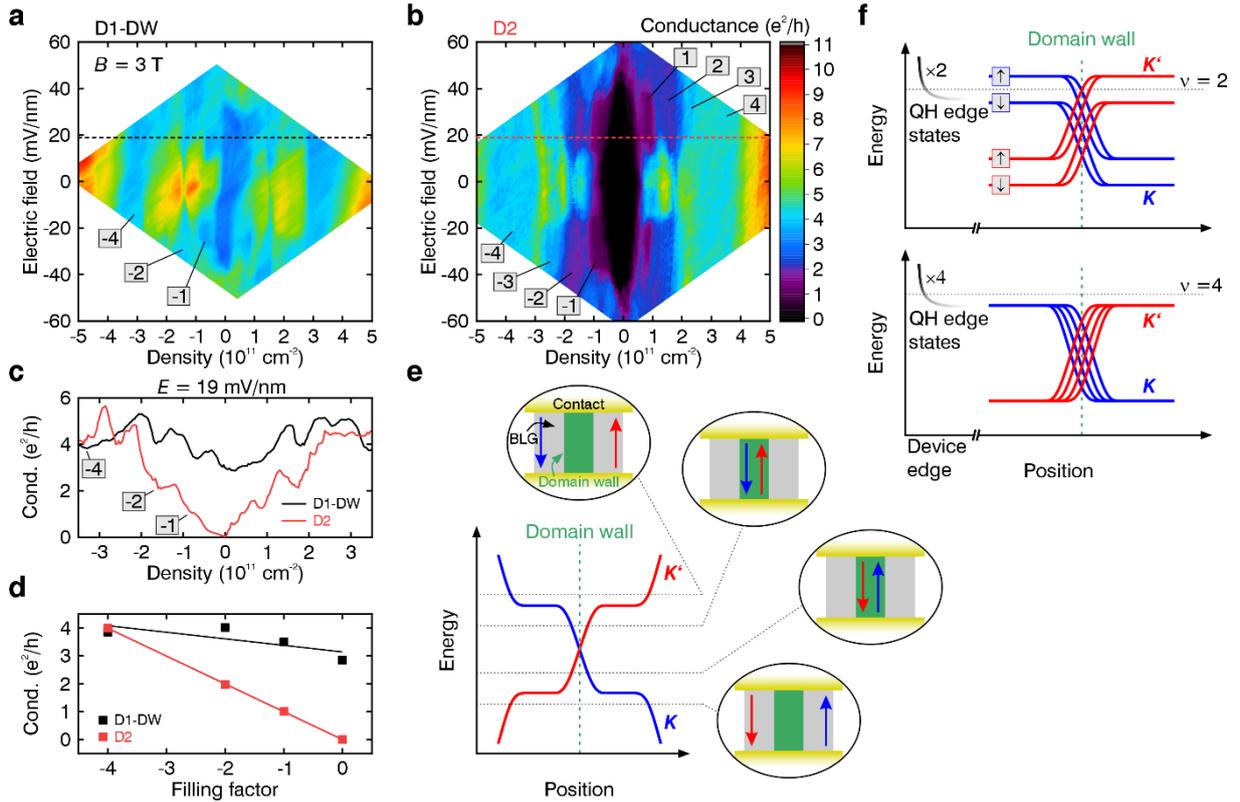

**Fig. 2 | Interplay between topological valley and quantum Hall edge transport at low magnetic fields. a,b,** Maps of the conductance in units of $e^2/h$ as a function of applied electric field $E$ and charge carrier density $n$ at a magnetic field of $B = 3$ T for devices D1-DW (a) and D2 (b). The dashed lines indicate the position of the data shown in (c). **c,** Line traces of the conductance as a function of $n$ taken at constant $E$ in device D1-DW (black) and D2 (red). **d,** Conductance of quantum Hall states as a function of filling factor for device D1-DW (black) and D2 (red). The values are averaged over the electric field range at which the individual states emerge. The solid lines are linear fits to the corresponding data. **e,** Schematic band structure (spin and orbital index omitted) in bilayer graphene in the presence of a stacking domain wall as a function of position. The dashed lines indicate distinct positions of the Fermi level and the encircled pictures schematically indicate the corresponding evolution of directions and locations of the one-dimensional channels within the device. **f,** Schematic band structure as a function of position across the device with a domain wall shown for the $\nu = 2$ (top) and $\nu = 4$ (bottom) QH state in the presence of an interlayer electric field (spin and orbital flavours have been reinstated).

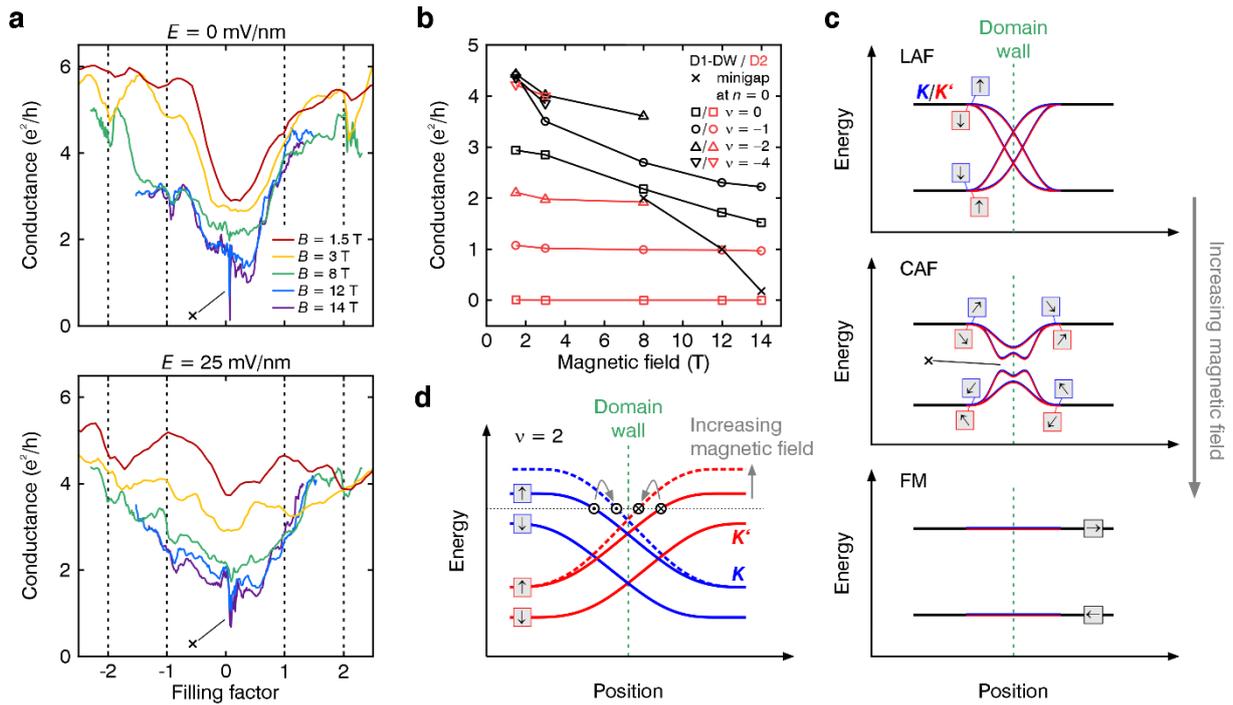

**Fig. 3 | Behaviour of the kink states for varying magnetic fields. a,** Conductance as a function of filling factor shown for various magnetic fields at $E = 0$ (top) and $E = 25$ mV/nm (bottom) measured in device D1-DW. The cross indicates the sharp conductance dip caused by the opening of minigap. **b,** Conductance of the $\nu = 0, -1, -2, -4$ quantum Hall states as well as within the minigap as a function of magnetic field. The data for device D1-DW (D2) is shown in black (red). **c,** Schematic band structure around the domain wall shown for the LAF, CAF and FM $\nu = 0$ phases. The blue (red) lines indicate the chiral states in the $K(K')$-valley. The cross indicates the spectral minigap emerging in the CAF phase. **d,** Schematic band structure for $\nu = 2$ (orbital index is implicit) in the presence of layer-polarising bias. The domain wall retains only two pairs of valley helical (spin polarised) states, indicated by black circles with in-plane and out-of-plane directions. Their backscattering rate at the chemical potential (thin horizontal line) depends on their spatial separation and width. Both are generally expected to change as a function of magnetic field, leading to a change in DW conductance. The influence of the magnetic field is indicated by grey arrows. A similar effect was observed in artificial domain walls[7].

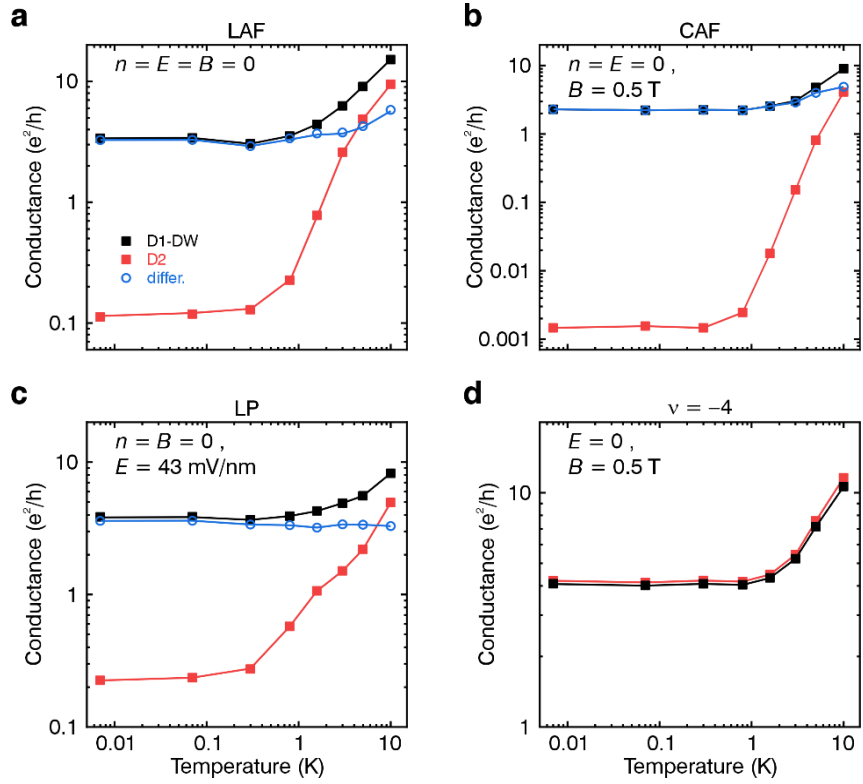

**Fig. 4 | Temperature dependence of the conductance in various broken-symmetry phases. a – d,** Temperature dependence of the conductance measured for the LAF phase at $n = E = B = 0$ (a), the CAF phase for $n = E = 0$ and $B = 0.5$ T (b), the LP phase at $n = B = 0$ and $E = 43$ mV/nm (c) and the $\nu = -4$ phase at $E = 0$ and $B = 0.5$ T. The data corresponding to device D1-DW (D2) is shown in black (red). Moreover, in panels (a – c), the difference of conductance between the two devices $\sigma_{diff}(T) = \sigma_{D1-DW}(T) - \sigma_{D2}(T)$ is shown as a function of temperature in blue. Note that the temperature dependence was measured in a different loading and annealing cycle than the measurements shown in Fig. 1 – 3 leading to small disparities in the conductance.

# Supplementary Information for

# Interplay between topological protected valley and quantum Hall edge transport

**Authors:** Fabian R. Geisenhof[1], Felix Winterer[1], Anna M. Seiler[1], Jakob Lenz[1], Ivar Martin[3], R. Thomas Weitz[1,2,4,5]*

**Affiliations:**

[1]Physics of Nanosystems, Department of Physics, Ludwig-Maximilians-Universität München, Geschwister-Scholl-Platz 1, Munich 80539, Germany

[2]1st Physical Institute, Faculty of Physics, University of Göttingen, Friedrich-Hund-Platz 1, Göttingen 37077, Germany

[3]Materials Science Division, Argonne National Laboratory, Lemont, Illinois 60439, USA

[4]Center for Nanoscience (CeNS), LMU Munich, Schellingstrasse 4, Munich 80799, Germany

[5]Munich Center for Quantum Science and Technology (MCQST), Schellingstrasse 4, Munich 80799, Germany

*Corresponding author. Email: thomas.weitz@uni-goettingen.de

**Identifying the emerging broken-symmetry quantum Hall states in the presence of a stacking domain wall**

Since the conductance of the appearing quantum Hall states in device D1-DW differs quite significantly from usually observed values as in device D2, we have additionally recorded fan diagrams to examine the slope of transconductance fluctuations[1,2] (see Fig. S1). As some of the emerging phases show an electric field dependence, we have measured the conductance at various applied fields (see Fig. S1a – d). The slopes of the appearing broken-symmetry states fit very well to the expected $\nu = 0, \pm 1, \pm 2$ and $\pm 4$ states, despite all having similar conductances of $3 - 4\ e^2/h$. Thus, additional charge transport along the domain wall in parallel to the quantum Hall edge states is unambiguously the cause for the higher conductance in the different states. Whereas the $\nu = 0$ phase is most stable for low and high electric fields, consistent with a phase transition from the LAF/CAF to the LP phase[3,4], the $\nu = \pm 4$ is most stable for low electric fields. Contrarily, the (partially) layer polarised $\nu = \pm 1, \pm 2$ phases appear only at intermediate electric fields at these low magnetic fields. Notably, since at $n = E = B = 0$ transconductance fluctuations with zero slope are visible, the layer antiferromagnetic phase is indeed present but masked owing to the quantum valley transport along the stacking domain wall.

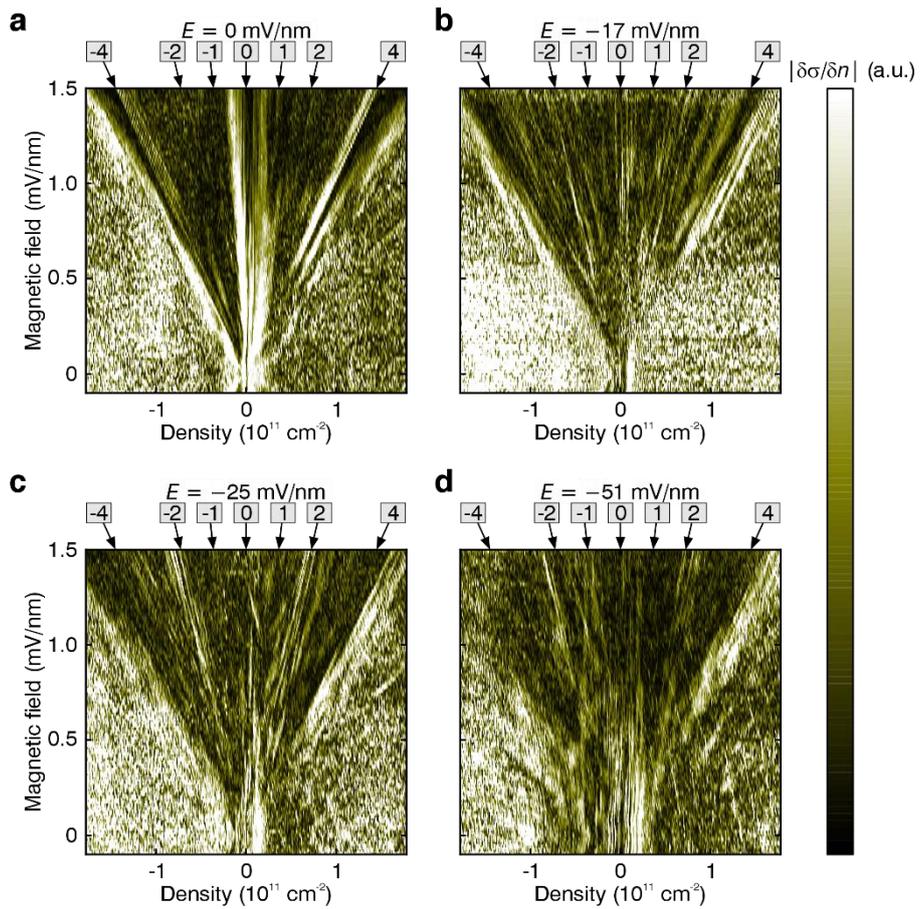

**Fig. S1 | Fan diagrams measured at specific electric fields. a – d,** Derivative of the differential conductance $\delta\sigma/\delta n$ as a function of magnetic field and charge carrier density for various electric fields. The slopes of the individual broken-symmetry $\nu = 0, \pm1, \pm2, \pm4$ states are indicated with arrows.

**Phase transition between the fully layer polarised and the canted antiferromagnetic $\nu = 0$ phase in the presence of a stacking domain wall**

When sweeping $E$ as a function of $B$ for zero charge carrier density (Fig. S2a,b), the transition between the fully layer polarised and the canted antiferromagnetic phase appears as region with increased conductance in both devices, consistent with previous measurements[3,4]. However, in device D2 both phases are insulating, whereas in D1-DW the conductance remains finite at $\sigma \approx 2.9\ e^2/h$ (see Fig. S2c) due to the kink states contributing to the charge transport. However, within the CAF phase the conductance is slightly decreasing for increasing magnetic field (Fig. S2d), consistent with the increase of canting and evolving energetic dispersion of the kink states (see main text). In Fig. S2e, the band structure of the CAF and LP phase in the presence of a domain wall are schematically shown. In contrast to the minigap opening in the CAF phase owing to the hybridizing of same valley states, the valley-helical kink states in the LP phase remain largely intact due to the suppression of intervalley scattering.

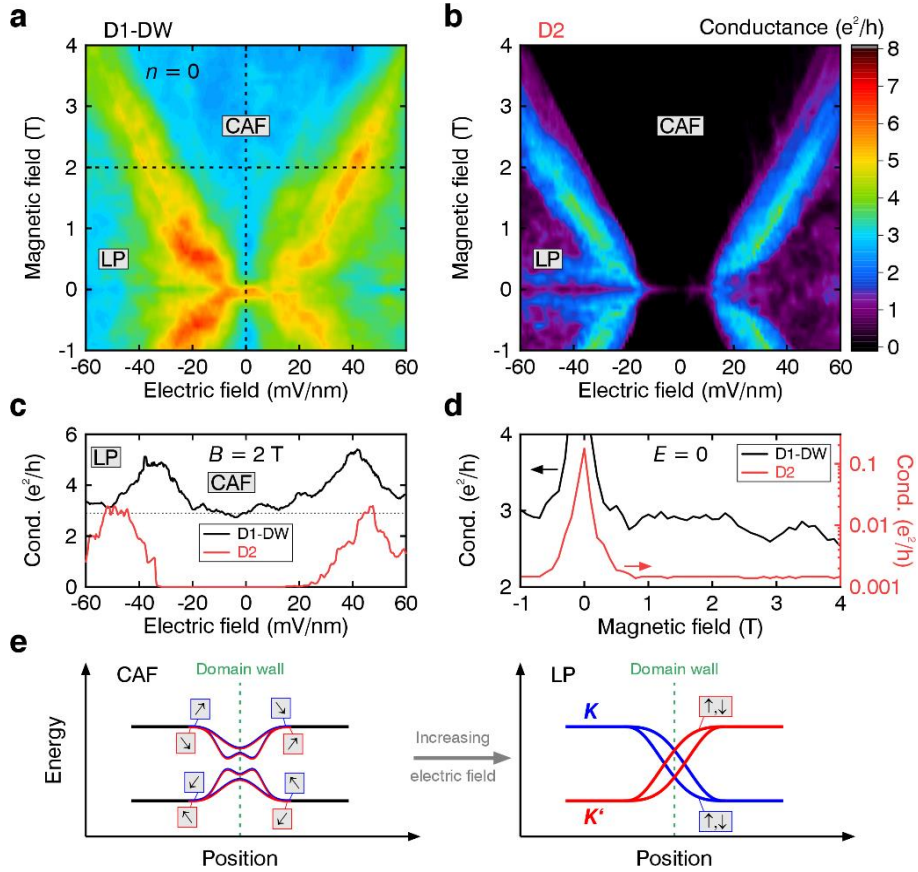

**Fig. S2 | Phase transition between the fully layer polarised and the canted antiferromagnetic $\nu = 0$ phase. a,b,** Conductance as a function of applied electric $E$ and magnetic field $B$ at zero charge carrier density for device D1-DW (a) and D2 (b). The first-order phase transition between the two $\nu = 0$ phases, the CAF and the LP phase, is characterized by a region with increased conductance. The dashed lines in (a) indicate the position of the data shown in (c) and (d). **c,** Line traces of the conductance across the $\nu = 0$ phase transition shown for sample D1-DW (black) and D2 (red) at $B$ = 2 T. The dashed line marks the value $2.9\ e^2/h$. **d,** Conductance as a function of $B$ for $n = E = 0$. The data of device D1-DW (D2) is shown in black (red). **e,** Schematic band structure as a function of position around a stacking domain wall shown for the CAF and LP phase.

**Impact of domain walls on the band structure beyond the zero energy Landau level**

Fig. S3 shows the influence of the domain wall on higher Landau levels. To this end, we have recorded a conductance map as a function of function of *E* and *n* at *B* = 1.5 T, see Fig. S3a. At this low magnetic field, we can observe the $\nu = \pm 8, \pm 12$ quantum Hall states. While the conductance of the $\nu = \pm 8$ states show the expected conductance, the one of the $\nu = \pm 12$ state seems to be lower than 12 $e^2/h$. Moreover, we see an oscillating behaviour of the conductance when entering a new quantum Hall plateau (see Fig. S3a,b). One possible explanation for these oscillations could be the rather low aspect ratio $L/W \approx 0.5$ of the device, with $L$ and $W$ being the length and width of the device channel, respectively. It has been shows that the shape of a sample can non-trivially affect conductance at the quantum Hall transitions[5].

However, theoretical calculations of Landau levels energies (see Fig. S3c,d) show that domain walls can cause the formation of ripples within higher Landau levels. For the band structure calculation, we used linearized model of graphene layers near $K/K'$ valley, with smoothly varying interlayer hybridization across the domain wall; namely, hopping A1B2 gradually is being replaced by B1A2. The energies are calculated in a finite width strip, with the domain wall in the middle. In Landau gauge, the translational invariance in the direction of the strip is preserved, and the energies are plotted as a function of momentum along the strip. The rescaled momentum also corresponds to the locations $x$ of the centres of individual states, when rescaled by $c/(eB)$, with $c$ being the speed of light. Interestingly, if the expected increase of A1A2 and B1B2 hybridization at the domain wall is not included, the zero energy Landau levels remain flat through the domain wall. In contrast, the higher Landau levels show significant variations – "ripples" – near the domain wall. This could also be the possible reason for the observed oscillations of conductance as a function of density.

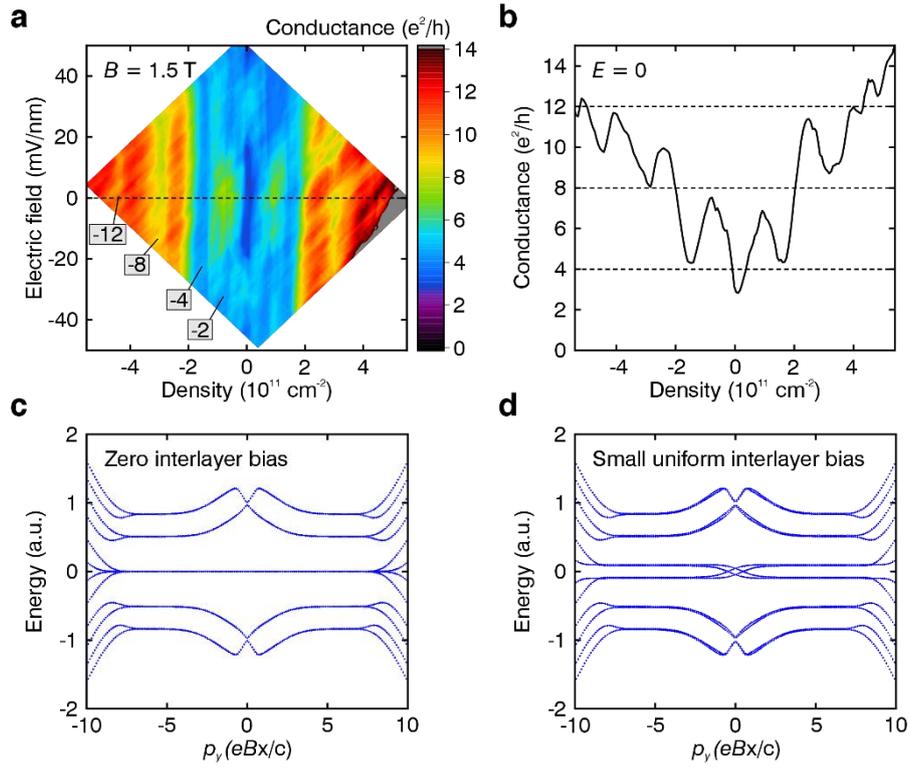

**Fig. S3 | Influence of the domain wall on the energetic landscape of higher Landau levels. a,** Conductance as a function of $E$ and $n$ at $B$ = 1.5 T. The dashed line indicates the position of the data shown in (b). **b,** Line trace of the conductance as a function of charge carrier density for zero electric field. The dashed lines indicate multiples of $4\,e^2/h$. **c,** Schematic Landau level band structure computed in the presence of a domain wall, smoothly interpolating between AB and BA stacking. The horizontal axis is the momentum along the domain wall. In Landau gauge used here, it is proportional to the location of the centre of a given orbital, $x$. With zero interlayer bias, the zeroth Landau level is four-fold degenerate (valley and orbital index), and higher Landau levels are doubly degenerate owing to valley degeneracy (spin is ignored). Near the edges the Landau levels float away from zero energy. The behaviour near the domain wall ($x = 0$) depends on the precise way that domain wall interpolates between AB and BA stackings. Even when the zeroth Landau level is flat (when A1A2 and B1B2 hopping near domain wall is ignored), the higher Landau levels are sensitive to the presence of the stacking defect. **d,** same as in (c) but with a small uniform interlayer bias. Notably, valley-helical modes emerge at the domain wall.

**Full quantum transport data at low and high magnetic fields in the presence of a stacking domain wall**

Fig. S4a – e shows the full conductance maps as a function of $E$ and $n$ for various magnetic fields in device D1-DW. Most prominently, the spectral minigap emerges for $B \geq 8$ T causing the conductance to drop, marked by the cross in Fig. S4c – e and in the line traces in Fig. S4f – h. Additionally, the conductances of the $\nu = 0, \pm 1, \pm 2$ states are dropping for increasing magnetic field. This can also be observed in the line traces shown in Fig. S4f – h. For the $\nu = 0$ CAF phase, the decrease can be explained by the emergence of a minigap due to the hybridizing of a partially spin aligned counterpropagating modes in the same valley, as explained in the main manuscript. For the $\nu = 0, \pm 1, \pm 2$ states, we think the effect occurs owing to increased intervalley scattering, as explained in the main manuscript.

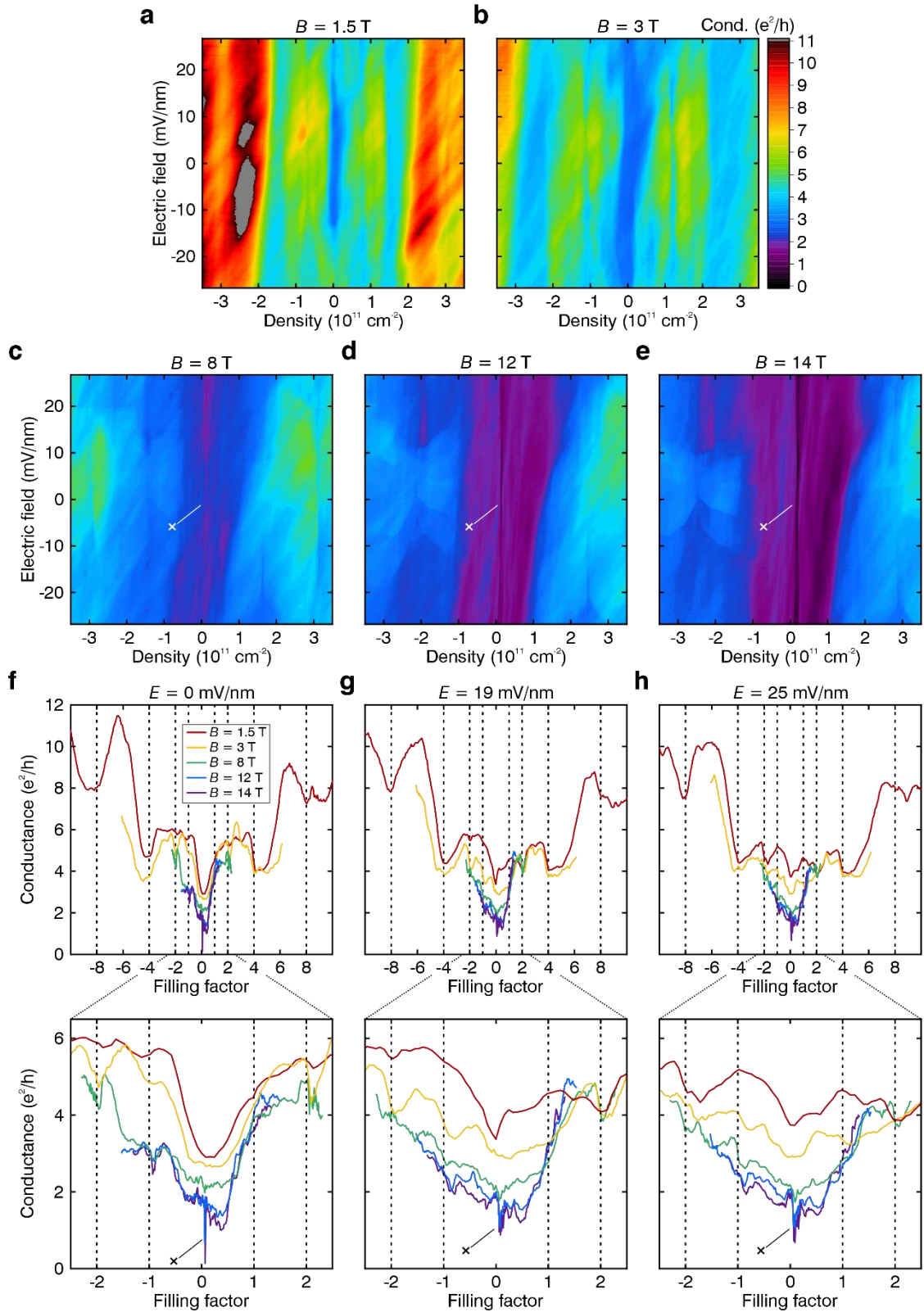

**Fig. S4 | Topological valley and quantum Hall edge transport for low and high magnetic fields. a – e,** Maps of the conductance as a function of electric field and charge carrier density for various magnetic fields in device D1-DW. The cross indicates the conductance dip caused by the appearing minigap. **f – h,** Line traces of the conductance as a function of filling factor for $E$ = 0, 19 mV/nm and 25 mV/nm. The lower panels are zoom-ins around small filling factors.

**Quantum transport measured in additional devices**

Fig. S5 shows the data from three additional devices with domain wall. Fig. S5a – c shows the conductance of the $\nu = 0, -1, -2, -4$ quantum Hall states as a function of magnetic field measured in the devices D2-DW, D3-DW and D4-DW, respectively. Note that the conductance was averaged over the regime at which the respective state emerges, i.e. for the $\nu = 0$ CAF phase around zero electric fields, for the $\nu = -1$ and $-2$ at $|E| \geq 10$ mV/nm and $|E| \geq 15$ mV/nm and the $\nu = -4$ state at all electric fields. The full conductance maps as a function of $E$ and $n$ for various magnetic fields are shown in Fig. S5d – e for the three devices.

Similar to device D1-DW, all three samples show a decrease of the conductance for the quantum Hall states with increasing magnetic field. Although in device D2-DW (Fig. S5a) and D3-DW (Fig. S5b) the decrease is very prominent, sample D4-DW (Fig. S5c) shows only a slight decrease of conductance. Moreover, a clear minigap can only be observed in device D2-DW (see Fig. S5a,d). However, we have indications that the quality of the devices D3-DW and D4-DW is significantly lower than that of D2-DW or even D1-DW. In device D4-DW, even at low magnetic fields, the conductances of the $\nu = 0, -2, -4$ states differ greatly. Evidently, the additional conductance originating from the kink states is highly reduced due to a low quality of the domain wall. In device D3-DW, the CAF phase can only be clearly observed for $B \geq 3$ T. Both observations indicate an overall lower quality of the two devices. Hence, the minigap can probably not be resolved due to disorder in both samples.

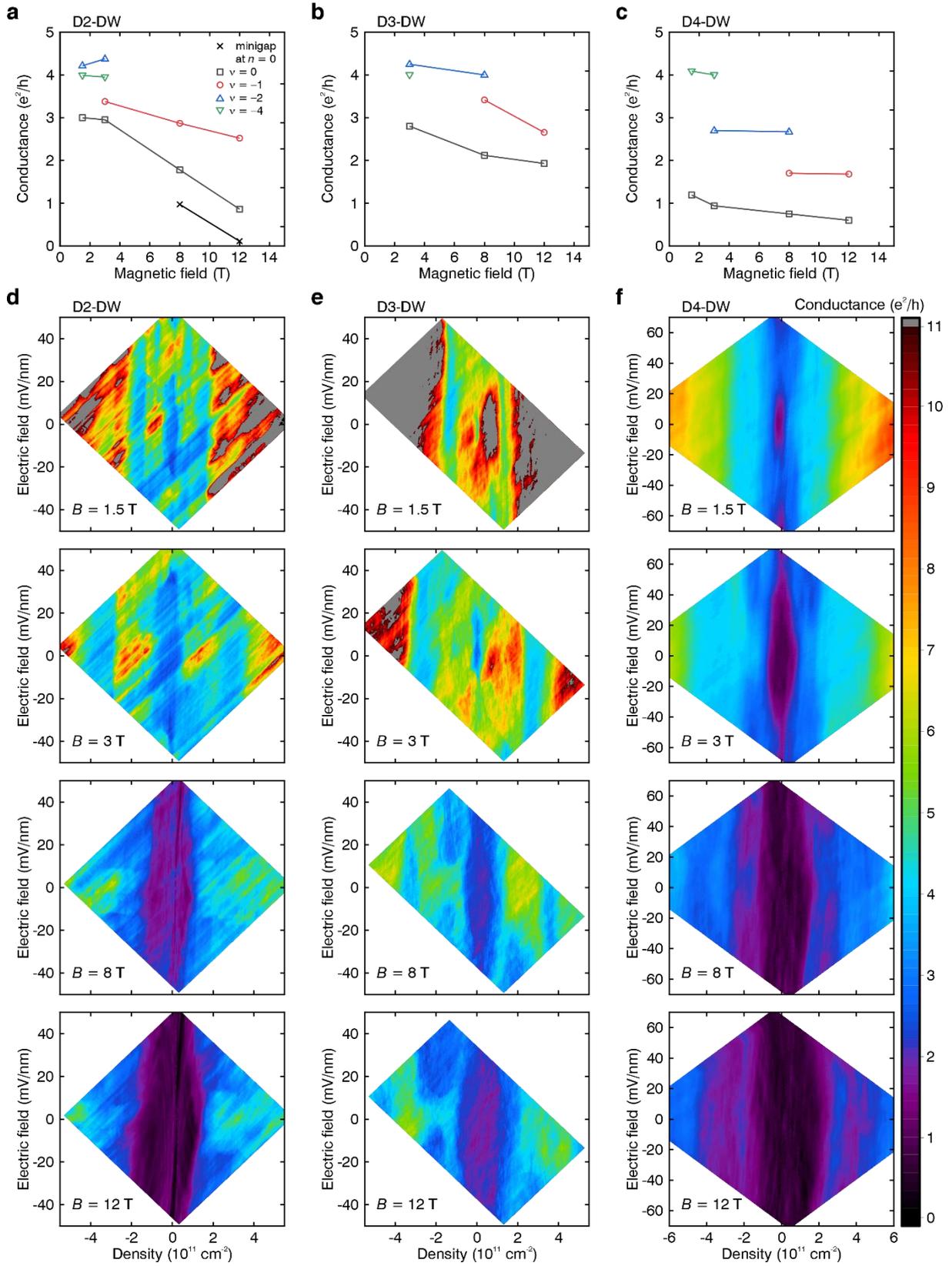

**Fig. S5 | Data from additional devices. a – c,** Conductance of the $\nu = 0, -1, -2, -4$ QH states as a function of $B$ for the devices D2-DW (a), D3-DW (b) and D4-DW (c). **d – f,** Maps of the conductance as a function of $E$ and $n$ for various magnetic fields for the three devices.